\documentclass[%
 reprint,
superscriptaddress,
 amsmath,amssymb,
 aps,
prb,
]{revtex4-2}

\usepackage{graphicx}% Include figure files
\usepackage{dcolumn}% Align table columns on decimal point
\usepackage{bm}% bold math
\usepackage{float}
\usepackage{amsmath}
\pdfoutput=1
\begin{document}
 
%\preprint{APS/123-QED}

\title{Triangular Kondo lattice in $\mathbf{YbV_6Sn_6}$ and its quantum critical behaviors in magnetic field}

\author{Kaizhen Guo}
%\altaffiliation[Also at ]{Physics Department, XYZ University.}%Lines break automatically or can be forced with \\
%\author{Shuang Jia}%
 %\email{Second.Author@institution.edu}
\affiliation{
International Center for Quantum Materials, School of Physics, Peking University, Beijing 100871, China
}%

\author{Junyao Ye}
%\altaffiliation[Also at ]{Physics Department, XYZ University.}%Lines break automatically or can be forced with \\
%\author{Shuang Jia}%
%\email{Second.Author@institution.edu}
\affiliation{
	International Center for Quantum Materials, School of Physics, Peking University, Beijing 100871, China
}%

\author{Shuyue Guan}
%\altaffiliation[Also at ]{Physics Department, XYZ University.}%Lines break automatically or can be forced with \\
%\author{Shuang Jia}%
%\email{Second.Author@institution.edu}
\affiliation{
	International Center for Quantum Materials, School of Physics, Peking University, Beijing 100871, China
}%

\author{Shuang Jia}
\homepage{gwljiashuang@pku.edu.cn}
\affiliation{
International Center for Quantum Materials, School of Physics, Peking University, Beijing 100871, China
}%
\affiliation{
 Interdisciplinary Institute of Light-Element Quantum Materials and Research Center for Light-Element Advanced Materials, Peking University, Beijing 100871, China
}%
\affiliation{
CAS Center for Excellence in Topological Quantum Computation,
University of Chinese Academy of Sciences, Beijing 100190, China
}%

\date{\today}% It is always \today, today,
             %  but any date may be explicitly specified

\begin{abstract}
	
We report magnetization, heat capacity and electrical resistivity for a newly discovered heavy fermion (HF) compound $\mathrm{YbV_6Sn_6}$ which is crystallized in a hexagonal $\mathrm{HfFe_6Ge_6}$-type structure, highlighted by the stacking of triangular ytterbium sublattice and kagome vanadium sublattice.
Above 2~K, $\mathrm{YbV_6Sn_6}$ shows typical HF properties due to the Kondo effect on the Kramers doublet of $\mathrm{Yb^{3+}}$ ions in crystal electric field.
A remarkable magnetic ordering occurs at~$T_{\mathrm{N}}=0.40$~K in zero field while a weak external field suppresses the ordering and induces non-Fermi liquid (NFL) behavior.
In strong magnetic field the compound shows a heavy Fermi liquid state.
$\mathrm{YbV_6Sn_6}$ presents as one of the few examples of Yb-based HF compounds hosting triangular Kondo lattice on which a magnetic field induces quantum criticality near zero temperature.

\end{abstract}

%\keywords{Suggested keywords}%Use showkeys class option if keyword
                              %display desired
\maketitle

%\tableofcontents

\section{INTRODUCTION}

The research on the heavy fermion (HF) compounds has been an important part in strongly correlated physics in the past few decades \cite{stewart1984heavy,RevModPhys.73.797,coleman2006heavy}.
As the relevant energy scales are small in the HF compounds, their ground states can be readily tuned, which provides convenience for the laboratory to induce quantum critical point (QCP) through pressure, magnetic field, and chemical substitution \cite{gegenwart2008quantum,paschen2021quantum}.
Quantum fluctuation dominates near QCPs, which emerges highly collective excitation and exotic quantum phases.  
%The competition between Kondo effect and Ruderman-Kittel-Kasuya-Yosida (RKKY) interaction determines the ground state of most HF compounds, and the low energy scales of these two effects provide convenience for the laboratory to induce quantum critical points (QCP) through pressure, magnetic field, and chemical substitution \cite{gegenwart2008quantum,paschen2021quantum}.
%Entropy accumulates near QCP \cite{PhysRevLett.118.107204}, which opens up the possibility of new phases and excitations.
The most striking is unconventional superconductivity near antiferromagnetic (AFM) QCP, which has been explored in many Ce- and U-based HF compounds \cite{yuan2003observation,ott1983u}.
Ferromagnetic (FM) QCP was reported in the stoichiometric $\mathrm{CeRh_6Ge_4}$ \cite{shen2020strange} as well.
Compared with Ce- and U-based compounds, Yb-based HF compounds are relatively rare.
The scarcity is due to the fact that Yb ions are more localized, or likely exist as non-magnetic divalent ions.
The large vapor pressure of Yb element makes it difficult to synthesize the compounds as well.
Unconventional superconductivity and non-Fermi liquid (NFL) behaviors were observed in $\mathrm{YbAlB_4}$ \cite{Nakatsuji2008} and $\mathrm{YbRh_2Si_2}$  \cite{schuberth2016emergence}.
Magnetic field can induced various quantum critical behaviors including NFL and multiple phase transitions in $\mathrm{YbPtBi}$ \cite{mun2013magnetic} and $\mathrm{YbAgGe}$ \cite{bud2004magnetic} . The studies on the Yb-based HF compounds have advanced the understanding of the NFL and Fermi surface reconstruction near the QCP \cite{RevModPhys.73.797,si2010heavy}.

In a classical Doniach phase diagram, the competition between Kondo effect and Ruderman-Kittel-Kasuya-Yosida (RKKY) interaction determines the ground state of the HF compounds \cite{doniach1977kondo}.
While the Doniach scenario predicts a single magnetic QCP between the AFM ordering and heavy Fermi liquid (FL) state, recent experimental studies revealed it is insufficient to cover the various quantum critical behaviors in many HF compounds \cite{vojta2018frustration}, because frustration effect plays an important role.
The studies on the HF compounds hosting geometric frustrated Kondo lattices, for example, Yb-based Shastry-Sutherland lattice in $\mathrm{Yb_2Pt_2Pb}$ \cite{kim2008yb,kim2011heavy,kim2013spin} and Ce-based distorted kagome lattice in $\mathrm{CePdAl}$ \cite{donni1996geometrically,zhang2018kondo,lucas2017entropy} and $\mathrm{CeRhSn}$ \cite{tokiwa2015characteristic}, have helped to understand a two-dimension global diagram determined by the Kondo effect and the strength of magnetic frustration \cite{si2006global,coleman2010frustration,si2014kondo}.
%The appearance of frustration can take various guises but the geometric frustrated Kondo lattices are mostly studied.

%The introduction of a Kondo effect ($J_K$) and strength of magnetic frustration ($G$) gives a global phase diagram for HF materials \cite{si2014kondo,coleman2010frustration,si2006global}.

%The real materials includes $\mathrm{Yb_2Pt_2Pb}$ \cite{kim2008yb}, containing Yb-based Shartry-Sutherland Kondo lattice, and $\mathrm{CePdAl}$ containing Ce-base distorted Kagome lattice \cite{donni1996geometrically}. 

Ytterbium-based compounds containing geometric frustrated triangular lattices are particularly interesting because the effective spin of the Kramers doublet of $\mathrm{Yb^{3+}}$ ion can be $J_\mathrm{eff} =1/2$.
These compounds provide a fertile ground for exploring exotic quantum matters such as valence bond solid (VBS) and quantum spin liquid (QSL) caused by the frustration-enhanced zero-point motions of spins \cite{coleman2010frustration, Balents2010}.
The Yb-based triangular lattices in insulating compounds $\mathrm{YbMgGaO_4}$ \cite{Paddison2017} and $\mathrm{NaYbO_2}$ \cite{Bordelon2019} were found to be platforms for exploring the QSL physics.
On the other hand, the inter-metallic compounds containing Yb-based triangular lattices, in which both the Kondo effect and geometric frustration play the role, have been less studied.
Theoretical studies on the triangular Kondo lattice have suggested multiple quantum states from various chiral-type magnetic ordering which may bring spontaneous Hall effect \cite{martin2008itinerant,kato2010stability}, to the partial Kondo screening (PKS) state in which a subset of moments forms Kondo singlet \cite{motome2010partial,noda2014partial}.
Discovery of Yb-based HF compounds with triangular Kondo lattice will provide an ideal playground for exploring the exotic states while the quantum control on the states will help to better understand the global phase diagram \cite{si2010quantum,PhysRevB.102.235125}.

In the studies on the $\mathrm{RMn_6Sn_6}$ family compounds \cite{Yin2020,PhysRevLett.126.246602}, we noticed that $\mathrm{HfFe_6Ge_6}$-type structure may serve as an ideal framework for bearing the Yb-based triangular Kondo lattice.
The `166' structure is highlighted by a kagome lattice of transition metal and a triangular lattice of rare earth.
Exploration of kagome materials in `166' family has motivated wide interest in strongly correlated and topological physics while previous focus was the novel topological properties of the flat band and Dirac band in the kagome lattice \cite{ishii2013ycr6ge6,li2021dirac,PhysRevLett.127.266401,xu2022topological}.
Although the rare earth elements play an important role in magnetic structure and topological properties \cite{PhysRevLett.126.246602}, the $4f$ electronic strong correlation is less reported in the grand `166' family.
The studies on  $\mathrm{YbMn_6Sn_6}$  and $\mathrm{YbMn_6Ge_6}$ demonstrated the valence change and magnetic ordering of Yb ions \cite{mazet2010valence,PhysRevLett.111.096402,PhysRevB.96.155129}.
There are few reports for the physical properties of other Yb-based `166' compounds \cite{Weiland2020, avila2005direct}.  

We successfully grew the single crystals of two new Yb-based `166' compounds, $\mathrm{YbV_6Sn_6}$ and $\mathrm{YbCr_6Ge_6}$ in the past two years.
While the characterization for the later will be presented somewhere else, we present the crystal structure, magnetization, heat capacity and electrical resistivity for $\mathrm{YbV_6Sn_6}$ in this study.
The existence of $\mathrm{RV_6Sn_6~(R = Y, Gd-Tm, Lu)}$ family was firstly reported by L.Romaka \textit{et.al.} in 2011 \cite{ROMAKA20118862}, but the physical properties were not reported until very recently.
In general, $\mathrm{RV_6Sn_6~(R = Gd-Tm)}$ demonstrate the physical properties of well-defined $4f$-local-moment bearing, weakly interacted magnetic intermetallic compounds \cite{PhysRevB.104.235139,lee2022anisotropic,zhang2022electronic,PhysRevB.106.115139,PhysRevMaterials.6.104202}.
Our measurements on $\mathrm{YbV_6Sn_6}$  above 2~K showed typical manners of HF compounds, including large electronic specific heat and coherent peak in temperature-dependent resistivity.
Moreover, we found a remarkable magnetic ordering at $0.40$~K in zero field while a weak magnetic field can suppress the ordering and induce NFL behavior.
In strong magnetic field the compound enters a heavy FL state with an electronic specific heat coefficient $\gamma > 400~\mathrm{mJ/mol~K^2}$.
We conclude that $\mathrm{YbV_6Sn_6}$ presents as a rare example of HF compounds which hosts Yb-based triangular Kondo lattice.

\section{EXPERIMENT Method}

Single crystals of $\mathrm{YbV_6Sn_6}$ and $\mathrm{LuV_6Sn_6}$ were synthesized via self-flux method. The starting elements of Yb(pieces, 99.9\%), Lu(pieces, 99.9\%), V(grains, 99.9\%) and Sn(shots, 99.97\%) with the molar ratio of R:V:Sn=1:6:20 were placed in an alumina Canfield Crucible Sets(CCS) \cite{canfield2016use}  which is effective to prevent samples from contacting the silica wool, and then sealed in a vacuum silica ampoule.
To avoid the slight reaction between Yb and alumina crucible, the small pieces of Yb and V were placed close together and surrounded by Sn drops.
The flux mixtures were heated to $\mathrm{1125^\circ C}$, dwell for 12 hours, then continuously cooled at $\mathrm{2^\circ C/h}$, and finally centrifuged at $\mathrm{780^\circ C}$ to remove excess flux.
The yielded crystals were hexagonal plates with the dimension $\mathrm {1\times 1\times 0.2~mm^3}$ (Inset in Fig.~\ref{fig1}~c).
Other $\mathrm{RV_6Sn_6~(R=Gd-Tm)}$ single crystals can be grown by this method as well.
The samples were soken into dilute hydrochloric acid for a short time to dissolve the excess tin flux on the surface.
Crystal structure was confirmed by measuring the powder X-ray diffraction (PXRD) at room temperature in a Rigaku Mini-flux 600 instrument.

Resistance and heat capacity measurements were carried out using a Quantum Design Physical Properties Measurement System (PPMS). A standard four-probe method was adapted in the resistance measurement with the current flowing perpendicular to the $c$ axis.
The heat capacity was only measured when $H\parallel c$ due to the limitation of crystals' size in this study.
%Because the samples are plate-like single crystals, a standing quartz base is needed to make sure the current flow is parallal to the $ab$-plane. However, only the $c$-axis direction was chosen to measure the specific heat.
A Dilution Refrigerator (DR) unit was used to measure low-temperature resistance and heat capacity from 0.2~K to 4~K.
%HERE YOU USE THE AC TRANSPORT OPTION IN PPMS? ME
To avoid sample heating, we use the $ac$ current $I = 0.2$~mA and $f = 33.6$~Hz.
Magnetization measurements was carried out using a Quantum Design Magnetic Properties Measurement Systems (MPMS-3) with a He-3 option.
The magnetization measurement revealed that the sample contains about 0.07 Vol. \% of tin impurity which contributes small diamagnetic signal below 4~K in a magnetic field less than 400~Oe.
To avoid misreading of the low-temperature magnetic susceptibility in weak fields, we subtracted the diamagnetic signal.  
%Single crystal samples were attached to brass rod and quartz rod using GE-Varnish. 

\section{RESULTS}

\subsection{Crystal Structure}

\begin{figure}[!htbp]
	\centering
	\includegraphics[width=\linewidth]{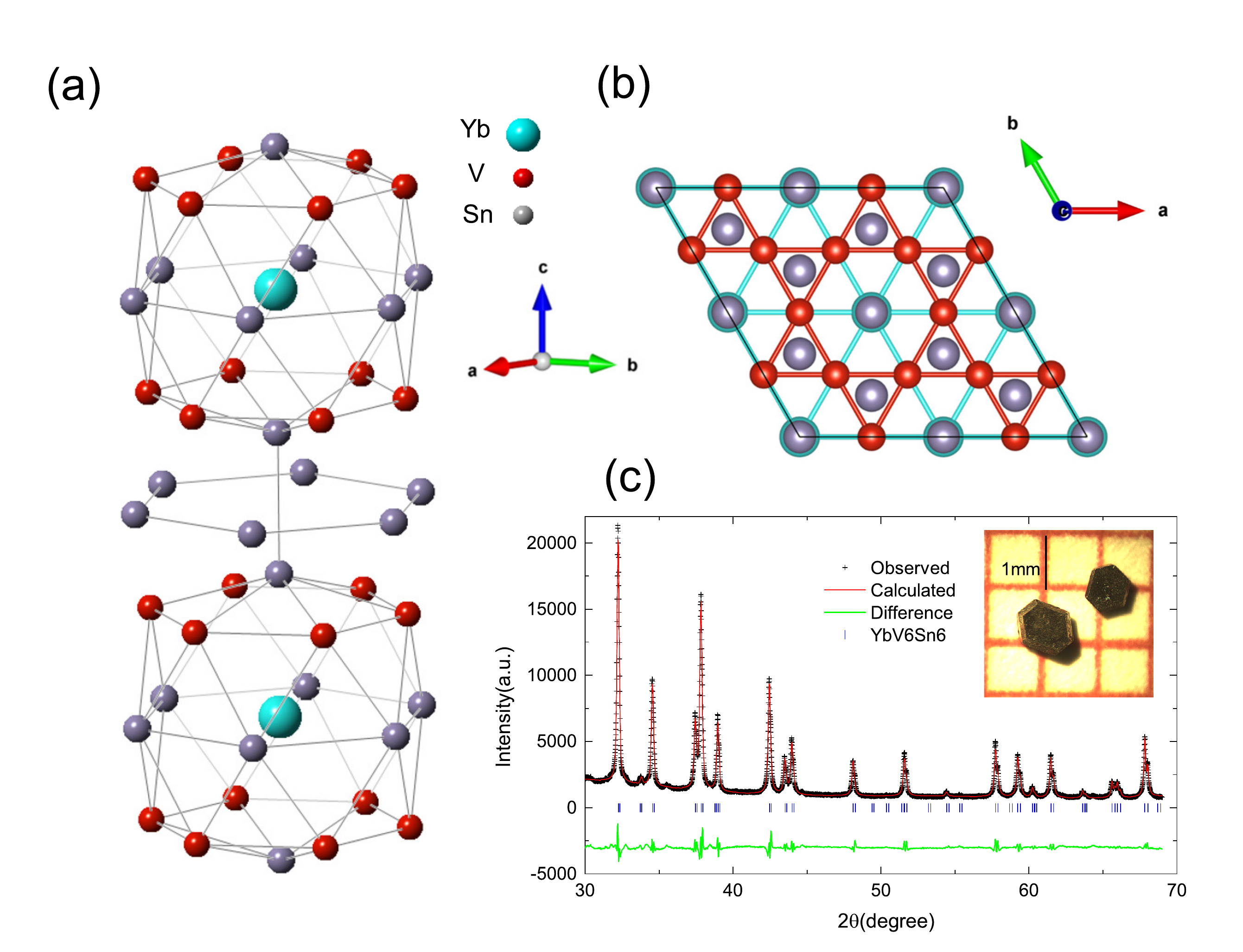}
	\caption{\label{fig1} (a) Crystal structure of $\mathrm{YbV_6Sn_6}$; (b) Triangular and kagome lattices formed by Yb and V atoms, respectively; (c) Measured (black dots) and calculated (red line) powder X-ray diffraction patterns. The vertical blue lines are calculated peak positions. Inset: a photograph of single crystals of $\mathrm{YbV_6Sn_6}$.}
\end{figure}

Most of the `166' compounds consisting of R, transition metal (T=V, Cr, Mn, Fe and Co) and germanium/tin atoms crystallize in  a $P6/mmm$ $\mathrm{HfFe_6Ge_6}$-type structure \cite{fredrickson2008origins} and so as $\mathrm{YbV_6Sn_6}$.
Highlighted by the pristine T-based kagome and R-based trianglular lattices  (Fig.~\ref{fig1}~b), the $\mathrm{HfFe_6Ge_6}$-type structure can be viewed as a R-stuffed CoSn-type structure, in which the R atom is caged in the polyhedron made by 12 T and 8 Sn atoms (Fig.~\ref{fig1}~a). 
The stuffed R atom pushes the Sn sites at the top and bottom of the void space away from the hexagonal center of V-based kagome net, leading to an alternation of stuffed and empty cavities along the $c$ axis.
Such arrangement leads to an extremely large nearest $\mathrm{Yb-Yb}$ distance ($c = 9.1701~ \mathrm{\AA}$) along the crystallographic $c$ axis while the nearest in-plane $\mathrm{Yb-Yb}$ distance of the triangular lattice is $a = 5.5020~ \mathrm{\AA}$.
The lattice parameters for $\mathrm{YbV_6Sn_6}$ are slightly larger than those for  $\mathrm{LuV_6Sn_6}$ ($a = 5.5016 ~\mathrm{\AA}$ and $c = 9.1692 ~\mathrm{\AA}$), indicating trivalent Yb ion at room temperature. 

We compare the parameters of the Yb-based triangular lattice to that in $\mathrm{YbAl_3C_3}$ ($a = 3.399~ \mathrm{\AA}$ and $c = 8.56~ \mathrm{\AA}$ ) \cite{gesing1992crystal,hara2012quantum} and $\mathrm{YbCuSi}$ ($a = 3.58~ \mathrm{\AA}$ and $c = 4.13~ \mathrm{\AA}$) \cite{iandelli1983low}.
In the distorted Yb kagome lattice in $\mathrm{YbAgGe}$, the nearest in-plane Yb-Yb distance is $3.664~ \mathrm{\AA}$ while the stacking distance is $4.14~ \mathrm{\AA}$ \cite{gibson1996ternary}.
Apparently $\mathrm{YbV_6Sn_6}$ has the longest stacking and relatively large in-plane distance.
We expect a weak, anisotropic RKKY interaction between the local moments associated with the $\mathrm{Yb^{3+}}$ ions.

\subsection{Heavy Fermion}

\begin{figure*}[!htbp]
	\centering
	\includegraphics[width=1\textwidth]{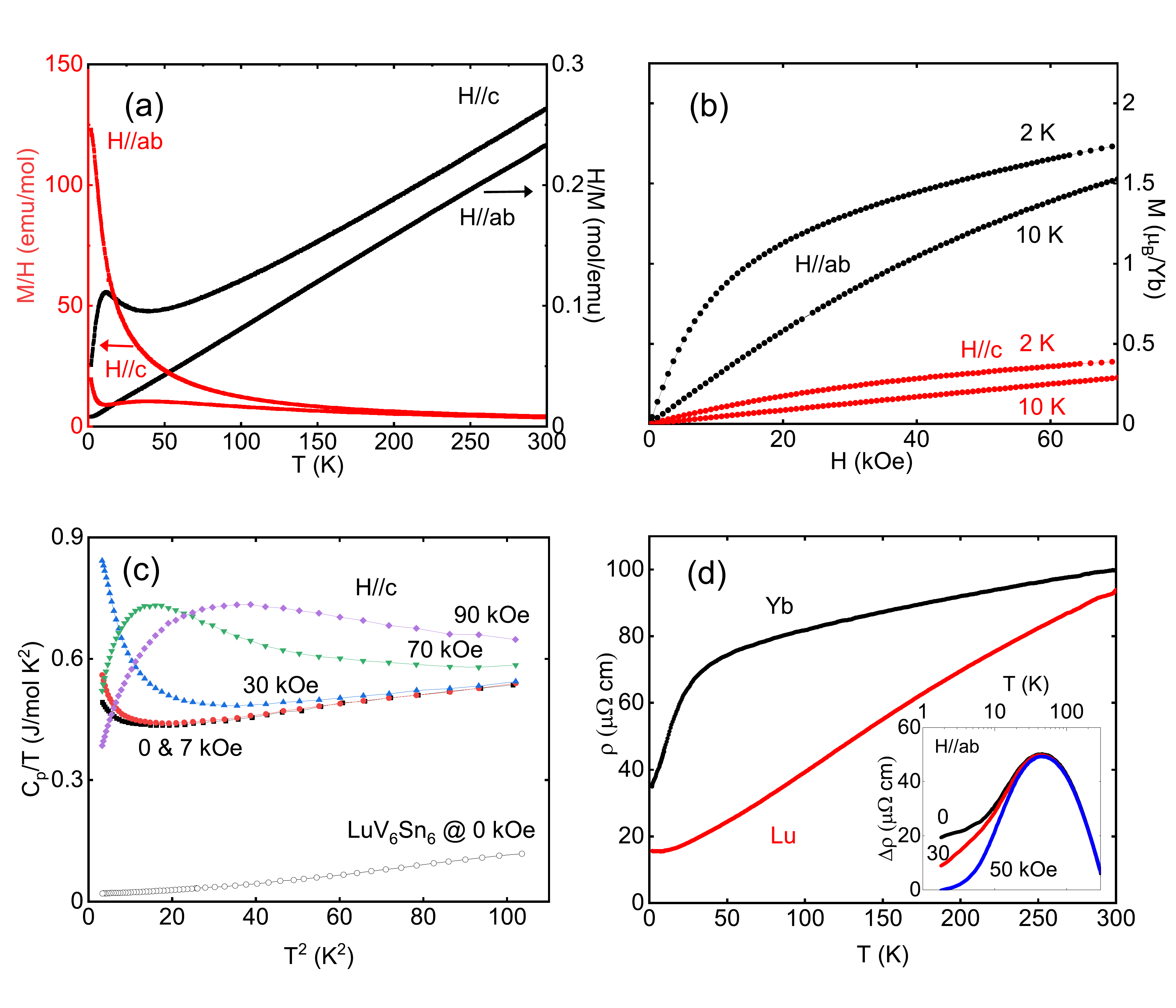}
	\caption{\label{fig2} Heavy fermions in $\mathrm{YbV_6Sn_6}$ above 2~K. (a) The magnetic susceptibility (red curve) and the inverse (black curve) in a magnetic field of $H = 30$~kOe for $H \parallel ab$ plane and $H \parallel c$ axis. (b) Field-dependent magnetization at 2 K and 10 K. (c) Low-temperature specific heat ($C_p$) divided by temperature, as a function of $T^2$ in different magnetic fields. $H$ was applied along the direction of $c$ axis. The bottom curve is the data for $\mathrm{LuV_6Sn_6}$ for comparison. (d) Temperature-dependent resitivity of $\mathrm{YbV_6Sn_6}$ and $\mathrm{LuV_6Sn_6}$ in zero field. Inset: $\Delta\rho$ vs $T$ for $\mathrm{YbV_6Sn_6}$ substracting the resitivity of $\mathrm{LuV_6Sn_6}$, from 1.8 K to 100 K, with the applied fields $H \parallel ab$ plane.}
\end{figure*}

Temperature-dependent magnetic susceptibility ($\chi (T)$) measurements of $\mathrm{YbV_6Sn_6}$ were carried out in a 30~kOe applied magnetic field $H\parallel c$ axis ($H_{\parallel c}$) and $H\parallel ab$ plane ($H_{\parallel ab}$) (Fig.~\ref{fig2}~a).
We notice a large easy-plane anisotropy at all temperatures.
When $H\parallel c$ axis, $\chi(T)$ shows a broad local maximum at about 45~K and then slightly drops to a local minimum at about 12~K and then increases again with temperature decreasing.
As comparison $\chi (T)$ for $H\parallel ab$ plane does not show any anomaly down to 2~K.
The profile of the anisotropic $\chi (T)$ for $\mathrm{YbV_6Sn_6}$ is reminiscent to that observed for $\mathrm{Yb_2Pt_2Pb}$ \cite{kim2008yb} and $\mathrm{CeCd_3P_3}$ \cite{lee2019two} and we believe it is due to crystalline electric field (CEF) effect.

The inverse magnetic susceptibility ($1/\chi(T)$) are linear and parallel to each other from 200~K to 300~K when the field is parallel to the directions of $c$ axis and $ab$ plane (Fig.~\ref{fig2}~a).
We fit this region by using the Curie-Weiss law, $\chi(T) = C/(T-\theta_{P})$, where $C=\frac{N_{A}\mu_\mathrm{eff}^2\mu_{B}^2}{3k_{B}}$ and $\theta_{P}$ is the Weiss temperature. Here $N_{A} = 6.02\times 10^{23}$  /mol is the Avogadro number, $\mu_{B}=9.274\times 10^{-24}$~J/T the Bohr magneton and $k_{B}=1.38\times 10^{-23}$~J/K the Boltzmann constant.
The results are  $\mu_\mathrm{eff}^{c} = 4.59~\mu_{B}$ and $\theta_{P}^{c} = -47.55~\mathrm{K}$ for $H\parallel c$ and  $\mu_\mathrm{eff}^{ab} = 4.54~\mu_{B}$ and $\theta_{P}^{ab} = -4.75$~K for $H\parallel ab$.
The obtained effective moment agrees well with the value for the Hund's rule ground state of $\mathrm{Yb^{3+}}$, $\mu_\mathrm{eff} = 4.54~\mu_{B}$, again proving that the Yb ion is trivalent at high temperature.
The large difference between $\theta_{P}^{c}$ and $\theta_{P}^{ab}$ reflects large anisotropy and we estimate the average Weiss temperature $\theta_{P}$ by $\theta_{P} = (\theta_{P}^{c} + 2\theta_{P}^{ab})/3=-19~\mathrm{K}$. 
The result is negative but much larger than the Weiss temperature for $\mathrm{GdV_6Sn_6}$ ($7.6$~K) in magnitude \cite{PhysRevB.104.235139}.
This large, negative Weiss temperature is comparable to many Yb-based HF compounds \cite{stewart1984heavy}.

Field-dependent magnetization $M(H)$ shows four time anisotropy for $H \parallel ab$ plane than $H \parallel c$ axis at 2~K (Fig.~\ref{fig2}~b).
For $H \parallel ab$, $M(H)$ increases linearly up to about 5~kOe, then shows a trend of saturation, and reaches $1.75~\mu_{B}$/Yb at 70 kOe, far less than the saturated moments of the Hund's rule ground state of $\mathrm{Yb^{3+}}~(4.0~\mu_{B})$.
%Below we will show the small magnetization in high field comes from the Kramer's doublet.
%The large anisotropy in the field dependent magnetization $M(H)$ at low temperature (Fig.\ref{fig2}b) is similar as observed in $\mathrm{YbAgGe}$ \cite{}.

Figure~\ref{fig2}~c shows the specific heat for $\mathrm{YbV_6Sn_6}$ and $\mathrm{LuV_6Sn_6}$ from 1.8~K to 10~K, presented as $C_{p}/T$ versus $T^2$.
As a non-magnetic counterpart, the specific heat for $\mathrm{LuV_6Sn_6}$ follows the relation of $C_{p}(T) = \gamma T + \beta T^3$ at low temperature where $\gamma$ equals $\mathrm{17.4~mJ/mol~K^2} $ and the $\beta$ value gives to the Debye temperature $\Theta_D$ being 150~K.
Assuming $\gamma$ is mainly comes from vanadium, we notice the value ($\mathrm{2.9~mJ/mol~Vanadium~K^2}$) is larger than that of regular metal such as gold and copper ($\mathrm{< 1~mJ/mol~K^2}$).

The low-temperature specific heat for $\mathrm{YbV_6Sn_6}$ at zero field follows the relation of $C_{p}(T) = \gamma T + \beta T^3$  from 3~K to 10~K, with the same value of $ \beta$ as in $\mathrm{LuV_6Sn_6}$.
 The $\gamma$ value, estimated as $\mathrm{411~ mJ/mol~K^2}$, is beyond the criteria for HF compounds ($>\mathrm{400~ mJ/mol~K^2}$) \cite{stewart1984heavy}.
Below 3~K,  $C_{p}/T$ shows a pronounced increase with temperature decreasing.
When a magnetic field is applied along the direction of $c$ axis, the low-temperature $C_{p}$ is significantly enhanced.
A broad peak occurs at around 4~K when $H=70$ kOe, and this peak shifts to higher temperature in stronger field.
The significant change of $C_{p}$ in magnetic field indicates that there remains a large part of magnetic entropy below 2~K which is released in a strong external field.
Below we show that the $\gamma$ value is indeed much larger than $\mathrm{400~mJ/mol~K^2}$ below 2~K.

We compare the temperature-dependent resistivity ($\rho(T)$) for $\mathrm{YbV_6Sn_6}$ and $\mathrm{LuV_6Sn_6}$ from 1.8 K to 300 K in Fig.~\ref{fig2}~d.
The $\rho(T)$ curve for $\mathrm{LuV_6Sn_6}$ behaves like a normal metal with a residue resistivity $\rho_0$ equaling $15.3 ~\mu\Omega~\mathrm{cm}$ and residual resistivity ratio $\mathrm{RRR = \rho(300 ~K)/\rho(2 ~K)}$ equaling 6.
The $\rho(T)$ curve for $\mathrm{YbV_6Sn_6}$, on the other hand, does not show a change of power law of $T^2$ at low temperature while the profile is commonly observed in many HF compounds.
We estimated the $4f$ part of the contribution by using  $\Delta\rho(T)=\rho_{\mathrm{Yb}}(T)-  \rho_{\mathrm{Lu}}(T)$ and the result is shown in the inset.
$\Delta\rho$ shows a broad maximum at $30$~K, signature of Kondo coherence and a $-log(T)$ dependence at higher temperature, incoherent Kondo scattering for it, respectively.
The resistivity is significantly suppressed in a magnetic field at low temperature but it changes little above 20~K.
The profiles of magnetization, heat capacity and resistivity unveil that $\mathrm{YbV_6Sn_6}$ can be viewed as a Kondo lattice of Yb ions, but it enters neither an FL state nor a magnetic ordering state above 2~K.

\subsection{Magnetic Transition}

\begin{figure}[!htbp]
	\includegraphics[width=\linewidth]{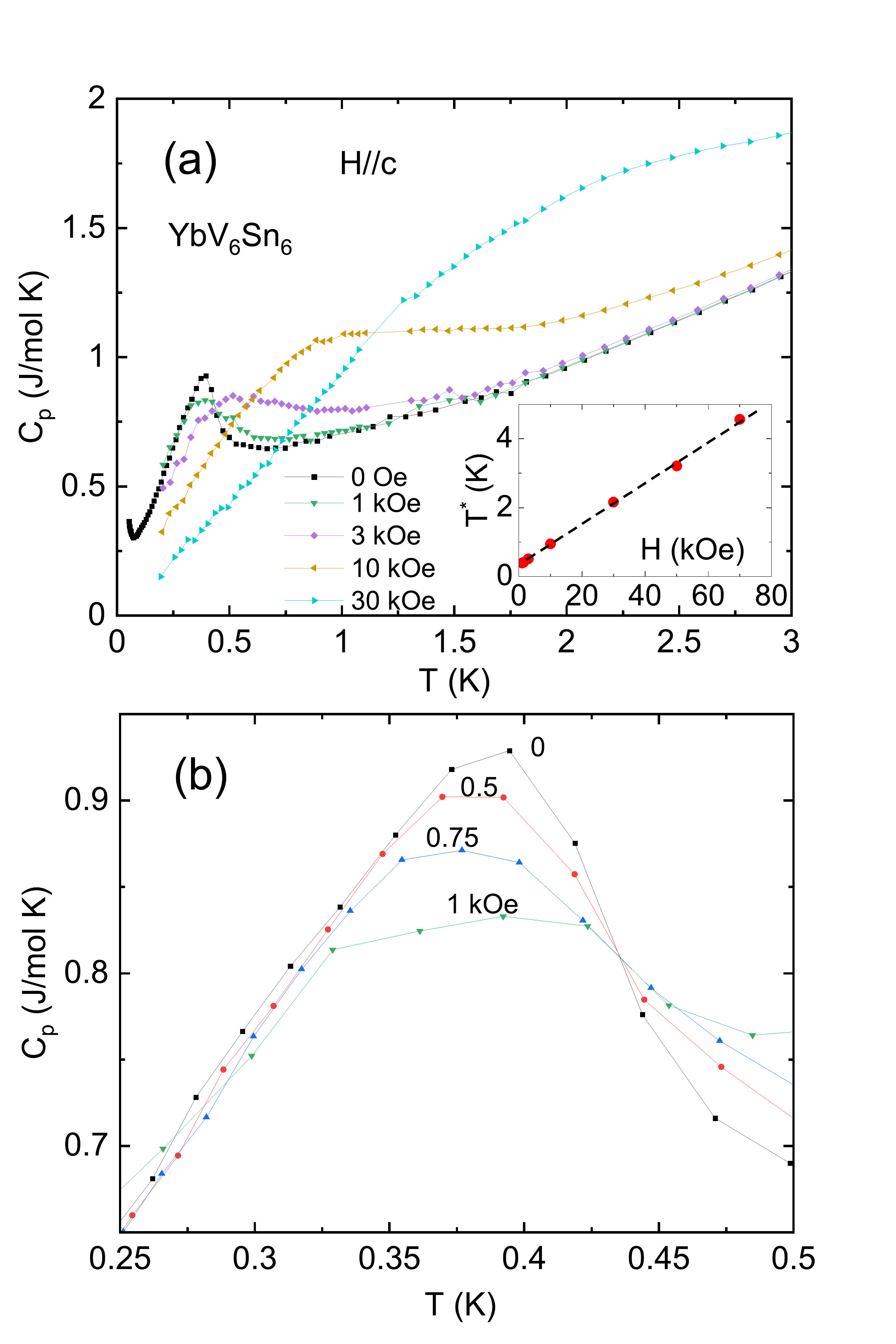}
	\caption{\label{fig3}Low-temperature specific heat for $\mathrm{YbV_6Sn_6}$. (a) Temperature dependent specific heat in different magnetic fields. The $\lambda$-shape peak at $0.4$~K in zero field becomes a broad peak in field. Inset: the peak position ($T^*$) with respect to the field.  (b) The $\lambda$-shape peak changes broad in weak magnetic fields. }
\end{figure}

To shed a light on the ground state of $\mathrm{YbV_6Sn_6}$  at low temperature, we measured the heat capacity and resistivity below 2~K by using a PPMS DR option.
The zero-field specific heat shows a sharp $\lambda$-shape peak at about 0.40~K, indicating a second-order phase transition on site.
When the temperature is below 0.1~K, $C_p$ shows an upturn due to a nuclear Schottky anomaly.
The $\lambda$-shape peak is suppressed by a magnetic field and a broad peak replaces it when $H_{\parallel c}=3~\mathrm{kOe} $.
In stronger field the peak shifts to higher temperature and finally becomes the feature we observed in Fig.~\ref{fig2}~c when $\mathrm {H\geq70~kOe}$.
As shown in the inset of Fig.~\ref{fig3}~a, the temperature of the $C_p$ maximum ($T^*$) shows a linear dependence on the field with a slope of $0.059$ K/kOe, indicating the broad peak is likely due to Zeeman effect.
If we consider the broad peak as a two-level Schottky anomaly, the peak position is satisfied with the relation of $k_BT^*=0.42E_Z$ where $E_Z=2MB$ is the Zeeman energy of the doublet in field \cite{lucas2017entropy}.
The magnetic moment estimated $M=0.70~\mu_{B}$ is close to the calculated  magnetization of a Kramers doublet ($M_{B\parallel z}=0.57~\mu_{B}$, see below).
 Moreover, the linear extrapolation of the $H$ dependence leads to an intercept field $H= -5.7$~kOe.
This negative intercept field implies an FM interaction which gives an additional effective field overlying on the external field. 

To check how the $\lambda$-shape peak evolves broad in weak field, we performed the measurements in the fields of 500~Oe, 750~Oe and 1000~Oe respectively (Fig.~\ref{fig3}~b).
As long as the peak shape is strongly modified by weak field, the peak center position only moves less than 0.02~K.
An interesting observation is that $C_p$ remains intact below $0.3$~K in the field weaker than 1000~Oe.
For a simple AFM ordered, local-moment-bearing material, the peak should be suppressed to 0~K gradually \cite{PhysRevB.61.9467}.
On the other hand, the specific heat change for $\mathrm{YbV_6Sn_6}$ in magnetic fields is similar with  HF compounds $\mathrm{Ce_2NiSi_3}$ \cite{szlawska2012magnetic} and $\mathrm{Ce_5Ni_2Si_3}$ \cite{lee2004magnetic}. The magnetic structures of these two HF compounds are sensitive to external magnetic field, and an FM alignment of magnetic moments can be easily induced by a weak field.

\begin{figure}[!htbp]
	\includegraphics[width=\linewidth]{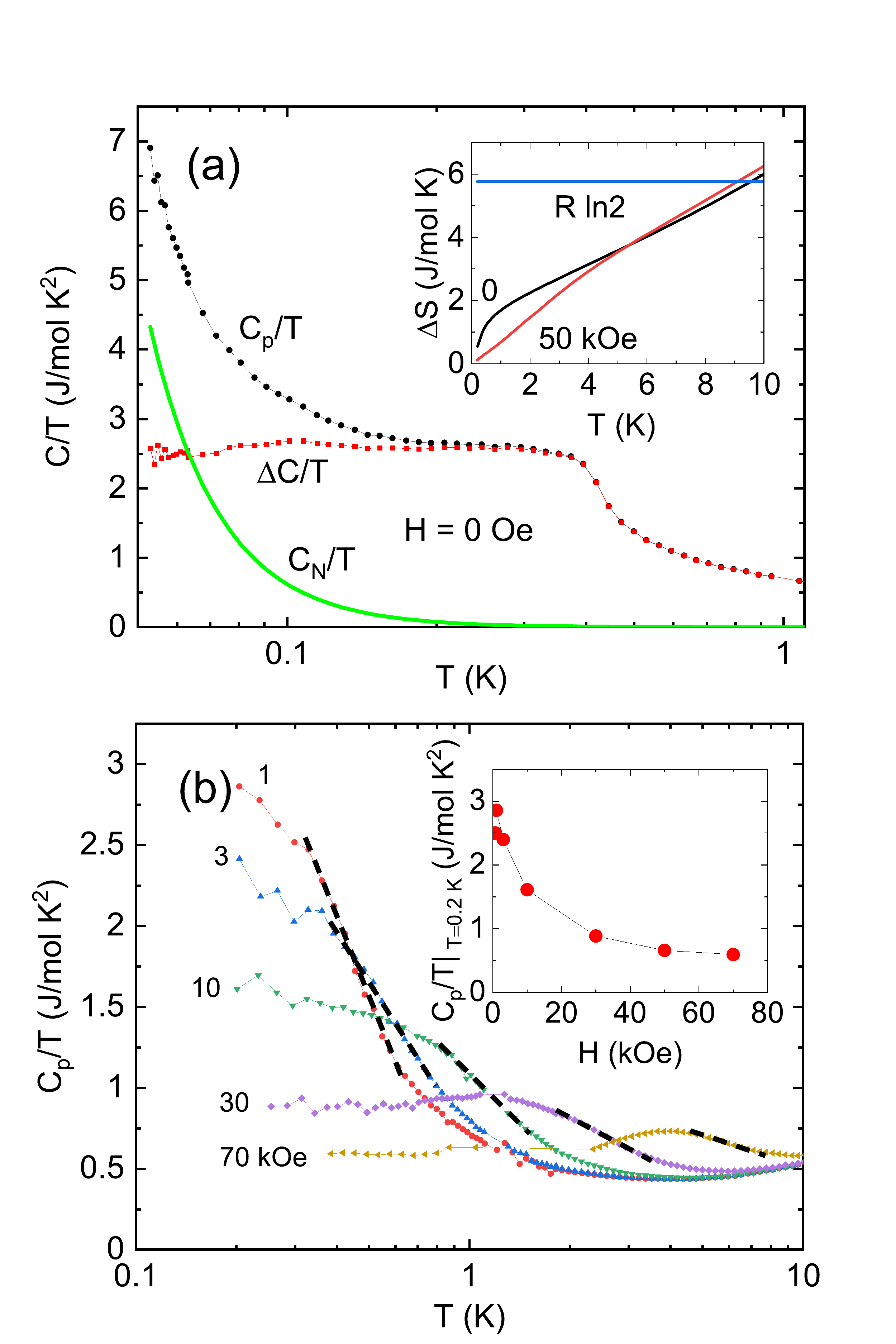}
	\caption{\label{fig4} Electronic specific heat of $\mathrm{YbV_6Sn_6}$. (a) Low-temperature $C/T$ in zero field, after subtracting the nuclear Schottky contribution. Inset: magnetic entropy $\Delta S_m$ below 10~K in $H=0$ and $50$~kOe. (b) $C_p/T$ versus $T$ in different magnetic fields. The dashed lines indicate the temperature region in which $C_p/T$ shows $-log(T)$ dependence.  Inset: $C_p/T$ at $0.2$~K as a function of field.} 
\end{figure}

We subtracted the nuclear Schottky contribution $C_N(T)$ which is proportional to $1/T^2$, from the $C_p$ below 2~K. While the phonon contribution can be ignored in low temperature, $\Delta C(T) /T$ is unchanged below $0.4$~K in zero field (Fig.~\ref{fig4}~a).
The low-temperature electronic specific heat coefficient $\gamma=\Delta C(T) /T|_{T = 0.2~\mathrm{K}}$ was estimated as $\mathrm{2.5~J/mol~K^2}$.
This value is much larger than that we estimated in the temperature range from 3~K to 10~K and comparable to that of $\mathrm{YbRh_2Si_2}$ \cite{PhysRevLett.89.056402}.

%This feature is similar as the observation in YbPtBi which shows magnetic ordering below $0.4$~K.
The magnetic entropy $S_m$  was inferred by means of integrating the $C_p/T$ when $H=0$ and $50$~kOe, after subtracting the $C_N(T)$ at low temperature and the $C_p/T$ of $\mathrm{LuV_6Sn_6}$ which is treated as the phonon contribution.
Noticing possible underestimation of $S_m$ below $0.05$~K that we cannot measure, we find that only $\sim 15\%$ of $R\ln(2)$ is released at $T_{\mathrm{N}}$, far less than the full magnetic entropy associated with a Kramers doublet.
The inferred $S_m$ recovers the full doublet entropy $R\ln(2)$ at 9~K, reaches $R\ln(4)$ at 28~K and $R\ln(8)$ at 45~K.
When $H= 50$~kOe, $S_m$ is released more slowly below 2~K but slightly faster above 2~K and recovers $R\ln(2)$ at 9~K as well.

Figure~\ref{fig4}~b shows the $C_p(T)/T$ in different $H_{\parallel c}$ from 0.2~K to 10~K.
Please note the nuclear Schottky anomaly and phonon contribution can be ignored in this temperature range.
When $H\leq 10$~kOe, $C_p(T)/T$ continuously increases with the temperature decreasing to 0.2~K and the value for $H=1$~kOe is even larger than the zero-field value.
The low-temperature value of $C_p(T)/T$ drops in a stronger field and becomes invariant with respect to temperature when $H\geq 30$~kOe,  indicating an FL ground state.
The inset shows the $C_p(T)/T$ values at 0.2~K with respect to the field.
The $\gamma$ values are much larger than $\mathrm{411~mJ/mol~K^2}$ when $H\geq 30$~kOe, again indicating that the electronic mass was seriously underestimated according to the data from 3~K to 10~K in zero field.

%For the specific heat data, $\Delta C_p(T)/T$ shows logarithmic [-ln(T)] behavior. Because $C_p(T)/T$ for $\mathrm{YbV_6Sn_6}$ is orders of magnitude larger than that in $\mathrm{LuV_6Sn_6}$, here we treat $C_p(T)/T$ as $\Delta C_p(T)/T$. $\Delta C_p(T)/T$ shows -ln(T) dependence between 0.2-0.5 K near 3 kOe and between 0.2-0.6 K near 10 kOe, with $H\parallel c$. $\Delta C_p(T)/T$ tend to be a constant below 0.7 K for H = 30 kOe, which is a Fermi liquid feature. For H = 5 kOe and H = 10 kOe, resitivity shows linear temperature dependence, revealing the deviation of $\mathrm{T^2}$ dependence, which is also seen as non-Fermi liquid signature. As a comparison, for H = 30 kOe, resitivity shows a $\mathrm{T^2}$ dependence below 1.1 K.
%The large part of the deficiency of the magnetic entropy is attributed to the Kondo screening effect or geometric frustration of the triangle lattice.
%Figure \ref{fig3}b shows the $\mathrm{C_p(T)}$ and the fitted $\mathrm{C_N(T)}$ and the $\mathrm{\Delta C(T)}$ at zero field, plotted as black circles, green line and red square, respectively.

\begin{figure}[!htbp]
	\includegraphics[width=\linewidth]{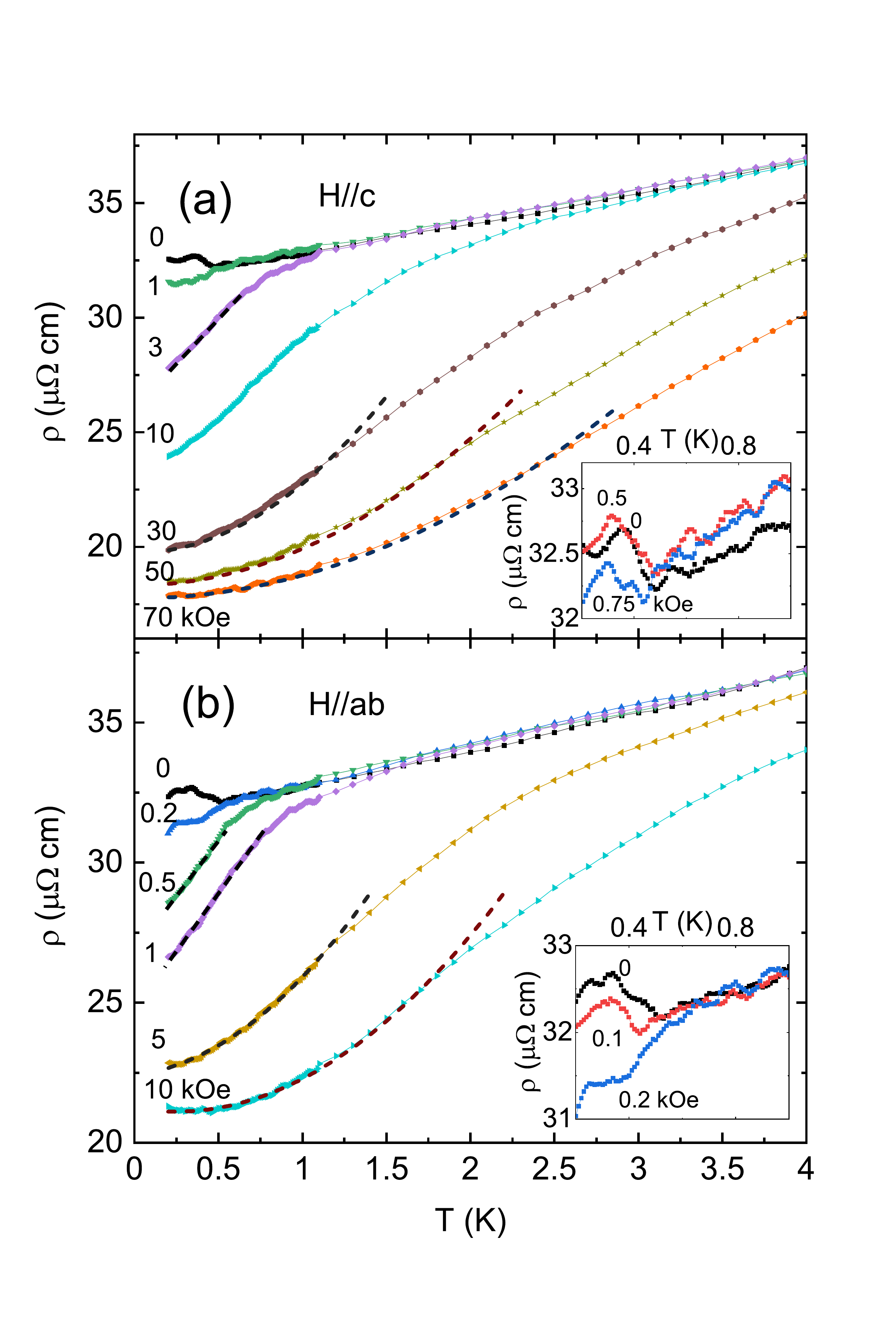}
	\caption{\label{fig5} Temperature-dependent resistivity in different magnetic fields. (a) $H\parallel ab$ plane. (b) $H\parallel c$ axis. The insets are the zoom-in in weak fields.}
\end{figure}

Figure~\ref{fig5}~a and b show $\rho(T)$ of $\mathrm{YbV_6Sn_6}$ when the field was applied along the directions of $c$ axis and $ab$ plane, respectively.
Zero-field $\rho(T)$ linearly decreases with the temperature decreasing to $0.4$~K and then shows a rise on site.
We compare the $\rho(T)$ for $\mathrm{GdV_6Sn_6}$ which has a magnetic ordering of well-defined local moments associated with $\mathrm{Gd^{3+}}$ ions at $5.0$ K \cite{PhysRevB.104.235139}.
The $\rho(T)$ of $\mathrm{GdV_6Sn_6}$ is similar with that of $\mathrm{LuV_6Sn_6}$ above 5~K and then it shows a slight drop below the ordering temperature, most likely caused by a loss of spin disorder scattering of the conduction electrons.
Indeed the $\rho(T)$ of $\mathrm{RV_6Sn_6~(R=Gd-Ho)}$ in general drops less than $0.5~\mu\Omega~cm$ from their magnetic ordering temperature to the based temperature \cite{lee2022anisotropic}. This $\rho(T)$ drop has much smaller magnitude than the $\rho(T)$ rise of $\mathrm{YbV_6Sn_6}$.
The profile of $\rho(T)$ for $\mathrm{YbV_6Sn_6}$ is reminiscent to that for YbPtBi \cite{mun2013magnetic}, a classical HF system which shows AFM ordering at $0.4$~K.
The rise of $\rho(T)$ is a general signature of charge density wave (CDW), spin density wave (SDW) transition and of AFM transition which opens a superzone gap.
It is noteworthy that our sample was mounted on the DR cold stage with GE varnish.
As shown in Ref.~\cite{mun2013magnetic}, mounting $\mathrm{YbPtBi}$ samples with GE varnish can bring local strain which weakens the rise of $\rho(T)$ during the transition, probably because the different thermal contraction between the sample and the cold stage brings stress.
Similar effect may exist in $\mathrm{YbV_6Sn_6}$ and future investigation is needed.

We notice $\rho(T)$ is very sensitive to the fields below $0.4$~K: $H_{\parallel ab}$ of 200 Oe and $H_{\parallel c}$ of 1~kOe can fully suppress the rise, respectively.
In moderate fields $H_{\parallel ab}$ of 1~kOe and $H_{\parallel c}$ of 3~kOe, respectively, the resistivity remains intact above 1~K but significantly drops at lower temperature.
This unusual manner leads to a `knee point' on $\rho(T)$ curve, below and above which $\rho(T)$ quasi-linearly increases with different slopes. 
The $\rho(T)$ curves show $T^2$ dependence at low temperature when $H_{\parallel ab}$ and $H_{\parallel c}$ is stronger than $5$ and $30$~kOe, respectively.
The residual resistivity ($\rho_0$) for $\mathrm{YbV_6Sn_6}$ is strongly suppressed by magnetic field, similar with those of HF compounds which are close to the magnetic instability \cite{flouquet1988residual}.

\subsection{Non-Fermi Liquid}

\begin{figure}[!htbp]
	\centering
	\includegraphics[width=\linewidth]{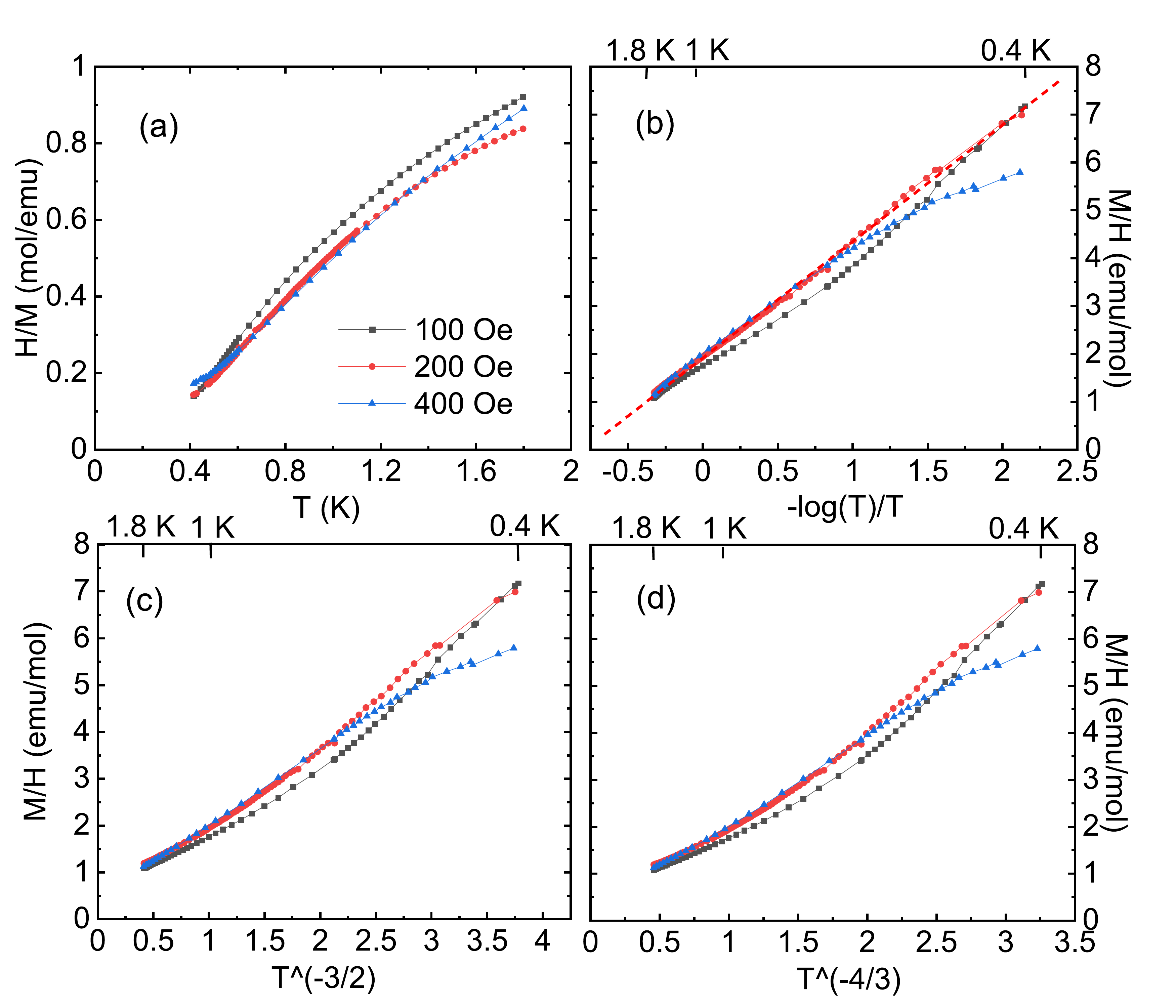}
	\caption{\label{fig6} Low temperature magnetization (a) $H/M$ as a function of $T$; (b), (c), (d) $M/H$ as a function of $-\log(T)/T$, $T^{-3/2}$ and $T^{-4/3}$, respectively.}
\end{figure}

%Therefore the kondo screening and the magnetic ordering barely affect the Kramer's doublet of Yb$^{3+}$.
When a moderate field suppresses the phase transition, we observed a linear power law  $\rho(T)\propto T$ at low temperature (Fig.~\ref{fig5}), which is a feature for an NFL.
As the knee points are below 1~K in general, the linear $\rho(T)$ only extends to very limited temperature.
This feature is unlike what observed in $\mathrm{YbRh_2Si_2}$ whose $\rho(T)$ linearly extends to about 20~K above the field-induce QCP \cite{PhysRevLett.85.626}.
On the other hand, $C_p(T)/T$ for $\mathrm{YbV_6Sn_6}$ only shows $-log(T)$ dependence in  limited temperature range (see dashed lines in Fig.~\ref{fig4}~b).
These observations demonstrate that $\mathrm{YbV_6Sn_6}$ can enter a quantum critical region of NFL when the transition is suppressed by a magnetic field.

To further elaborate the transition and field-induced quantum criticality, we performed $dc$ magnetization measurement from $1.8$ to $0.4$~K (field-cooling) when $H\parallel ab$ plane.
At the lowest temperature we accessed ($0.4$~K), the magnetic susceptibility ($\chi(T)=M/H$) is apparently deviated from the CW law  (Fig.~\ref{fig6}~a), indicating a nearby magnetic transition.
As plotting the $M/H$ with respect to $\log(T)/T$, $T^{-3/2}$ and $T^{-4/3}$ when $H$ is 100, 200 and 400~Oe, we notice that $\chi$ is linearly dependent on $\log(T)/T$ (Fig.~\ref{fig7}~b) in 200~Oe, which is the critical field suppress the transition exactly, according to the $\rho(T)$ measurement.
The dependence of $-\log(T)/T$ for $\chi$ is consistent with Moriya's model for the QCP in AFM 2D systems \cite{RevModPhys.73.797}. 
The power laws for the QCP in AFM 3D ($\chi \propto T^{-3/2}$) and FM 3D systems ($\chi \propto T^{-4/3}$) do not cover our observation (Fig.~\ref{fig7}~c and d).

\section{Analysis and discussion}

\begin{figure}[!htbp]
	\centering
	\includegraphics[width=\linewidth]{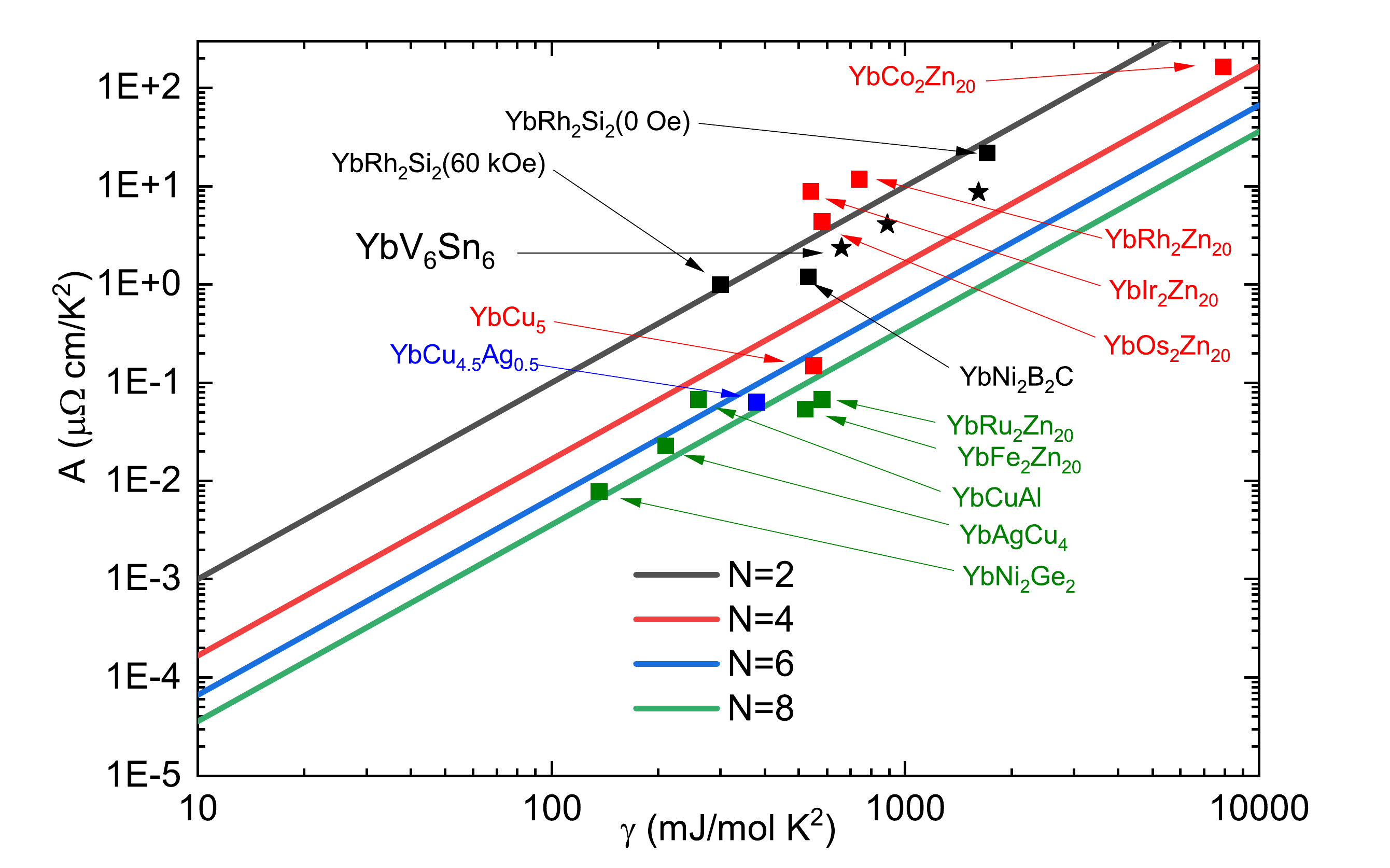}
	\caption{\label{fig7} Log-log plot of $A$ vs $\gamma$. Three sets of $\mathrm{YbV_6Sn_6}$ data for $H=10,30$ and $50$~kOe are represented as the stars from upper-left to bottom-right, respectively. The values for other representative Yb-based compounds are from Ref. \cite{PhysRevLett.85.626,PhysRevLett.89.056402,PhysRevB.55.1032,PhysRevB.56.8103,PhysRevB.54.R3772,fisk1992existence,Torikachvili9960} and the references therein.}
\end{figure}

The $4f$ part of the specific heat for $\mathrm{YbV_6Sn_6}$ recovers the full doublet entropy $R\ln(2)$ at 9~K, reaches $R\ln(4)$ at 28~K and $R\ln(8)$ at 45~K, which gives us an estimation of the energy splitting for the Hund's rule ground state of $\mathrm{Yb^{3+}}$ in CEF.
The CEF also leads the an-isotropic $\chi(T)$ which has a similar profile as the $\chi(T)$ for $\mathrm{TmV_6Sn_6}$ and $\mathrm{ErV_6Sn_6}$ \cite{lee2022anisotropic,zhang2022electronic}.
As the $\mathrm{Yb^{3+}}$ is located in a hexagonal coordination, the CEF Hamiltonian is given as
$\begin{aligned}
    \mathcal{H}_{\mathrm{CEF}}=B_{2}^{0} O_{2}^{0}+B_{4}^{0} O_{4}^{0}+B_{6}^{0} O_{6}^{0}+B_{6}^{6} O_{6}^{6},
\end{aligned}$
where $B_n^m$ are CEF parameters and $O_n^m$ are Steven operators \cite{hutchings1964point}.
The existence of $B_{6}^{6}$ leads to a mix between $\left|\pm \frac{7}{2}\right\rangle$ and $\left|\mp \frac{5}{2}\right\rangle$, thus gives rise to two new doublets. 

$$
\begin{aligned}
	\left|\Gamma_{\text {mix }, 1}\right\rangle &=\cos \alpha|\pm 7 / 2\rangle+\sin \alpha|\mp 5 / 2\rangle, \\
	\left|\Gamma_{\text {mix }, 2}\right\rangle &=\sin \alpha|\pm 7 / 2\rangle-\cos \alpha|\mp 5 / 2\rangle, \\
	\left|\Gamma_{3 / 2}\right\rangle &=|\pm 3 / 2\rangle, \left|\Gamma_{1 / 2}\right\rangle =|\pm 1 / 2\rangle
\end{aligned}
$$

From the Weiss temperatures difference for $H\parallel c$ axis and $H\parallel ab$ plane, we infer the CEF parameter $B_{2}^{0}$ being 2.37~K which is close to the estimation for $\mathrm{TmV_6Sn_6}$ \cite{lee2022anisotropic}.
%Unfortunately we cannot determine the other CEF parameters by fitting the $\chi(T)$ data.
The Kramers doublet $\left|\Gamma_{1 / 2}\right\rangle =|\pm 1 / 2\rangle$ is believed to be the CEF ground state, because the calculated saturated magnetization parallel and perpendicular to the $c$ axis are:
$$
\begin{aligned}
	&M_{B \parallel z}=\left\langle\Gamma_{\frac{1}{2}, 1}\left|J_{z}\right| \Gamma_{\frac{1}{2}, 1}\right\rangle g_{J} \mu_{B}=0.5(8 / 7) \mu_{B}=0.57 \mu_{B} \\
	&M_{B \perp z}=\left\langle\Gamma_{\frac{1}{2}, 1}\left|J_{x}\right| \Gamma_{\frac{1}{2}, 1}\right\rangle g_{J} \mu_{B}=2(8 / 7) \mu_{B}=2.29 \mu_{B}.
\end{aligned}
$$
The calculated $M_{B \parallel z}$ is a quarter of $M_{B \perp z}$, which is consistent with the  anisotropy of $M(H)$ curves at 2~K (Fig.~\ref{fig2}~b).
The value of $M_{B \parallel z}$ is close to $M$ at 2~K in 70~kOe ($0.4~\mu_{B}$/Yb) as well.
Thus $\mathrm{YbV_6Sn_6}$ hosts a triangular Kondo lattice in which the effective spin of the Kramers doublet is $J_\mathrm{eff} =1/2$.

We now try to understand the ground state of $\mathrm{YbV_6Sn_6}$  in variance of magnetic fields. When $H_{\parallel c}\geq30$~kOe and $H_{\parallel ab}\geq6$~kOe, respectively, the temperature-independent $C_p/T$ and the $T^2$-dependent $\rho(T)$ explicitly point to an FL ground state at low temperature. 
We obtained the coefficient $A$ by fitting the data of $\rho(T)=\rho_0+AT^2$ in the FL state and then plotted it as a function of $\gamma$ when $H_{\parallel c}= 30,~50$ and $70$ kOe.
The plot, known as Kadowaki-Woods (KW) plot \cite{kadowaki1986universal} (Fig.~\ref{fig7}), has includes other Yb-based HF compounds which have different degenerated ground states.
The points for $\mathrm{YbV_6Sn_6}$ are close to the line of $\mathrm{N = 2}$, implying that the Kondo screening effects on the Kramers doublet state for $\mathrm{Yb^{3+}}$, consistent with our analysis of the CEF.
Heaviness of the effective electronic mass of $\mathrm{YbV_6Sn_6}$ is highlighted although it has been strongly suppressed by the magnetic field.

As we are lack of the magnetization measurement below $0.40$~K at this point, the nature of the phase transition in zero field is not clear.
Noticing the zero-field $\rho(T)$ shows a rise on site, similar as what observed in $\mathrm{YbPtBi}$ \cite{mun2013magnetic}, it is plausible that there exists an SDW-type AFM ordering which may partially gap the Fermi surface, but other exotic phase transition cannot be ruled out. Below we envisage possible ground states for $\mathrm{YbV_6Sn_6}$.

The dilute R concentration and the long $\mathrm{R-R}$ distances are highlighted in the `166' family, and we expect weak RKKY interaction and low magnetic ordering temperature for R moments.
For the dilute Yb-based HF compounds such as $\mathrm{YbT_2Zn_{20}~(T = Fe, Co, Ru, Rh, Os, Ir)}$, the Kondo effect in general prevails and leads to a heavy FL ground state with no magnetic ordering down to 20~mK \cite{Torikachvili9960}. 
As other $\mathrm{RV_6Sn_6}$ compounds show extremely low magnetic ordering temperatures \cite{lee2022anisotropic,zhang2022electronic}, we infer the RKKY interaction in $\mathrm{YbV_6Sn_6}$ is very weak, because Yb has the smallest de Gennes factor ($dG = 0.32$) which scales the strength of RKKY interaction. If the de Gennes' law plays a role for the magnetic ordering in $\mathrm{RV_6Sn_6}$, the $T_{\mathrm{N}}$ for $\mathrm{YbV_6Sn_6}$ shall be below 0.1~K because $\mathrm{GdV_6Sn_6}$ has $T_{\mathrm{N}}=5.0$~K and $dG=15.75$.
Actually recent studies showed that the $T_{\mathrm{N}}$s for $\mathrm{R=Gd-Ho}$ approximately follow the de Gennes' law while $\mathrm{ErV_6Sn_6}$ and $\mathrm{TmV_6Sn_6}$ show no magnetic ordering down to 0.4~K \cite{lee2022anisotropic,zhang2022electronic}.
The unexpected $T_{\mathrm{N}}=0.40$~K is extraordinarily high in $\mathrm{YbV_6Sn_6}$ because both Kondo effect and geometric frustration in a triangular lattice shall impede the magnetic ordering.

Theoretical studies suggested that a 2D triangular Kondo lattice can have various emergent quantum states, from chiral magnetic ordering of the local moments \cite{akagi2010spin} to a PKS state \cite{motome2010partial,noda2014partial}. 
The PKS states have been found experimentally in the Ce-based distorted kagome lattice in $\mathrm{CePdAl}$ \cite{donni1996geometrically} and U-based triangular lattice in $\mathrm{UNi_4B}$ \cite{mentink1994magnetic}.
In a PKS state, the 2D triangular Kondo lattice resolves the frustration by means of forming a subset of non-magnetic Kondo singlet sites while the remaining magnetic sites forming an AFM honeycomb lattice.
If similar PKS exists in $\mathrm{YbV_6Sn_6}$ in zero field, the magnetic ordering may be largely different from that in other $\mathrm{RV_6Sn_6}$ and the breaking of the de Gennes' law is not unexpected.

%It is not clear whether this transition can be suppressed to 0~K, but we do not observe any evidence of the transition temperature changing in weak fields parallel to the $c$ axis.
%

\begin{figure}[!htbp]
	\centering
	\includegraphics[width=\linewidth]{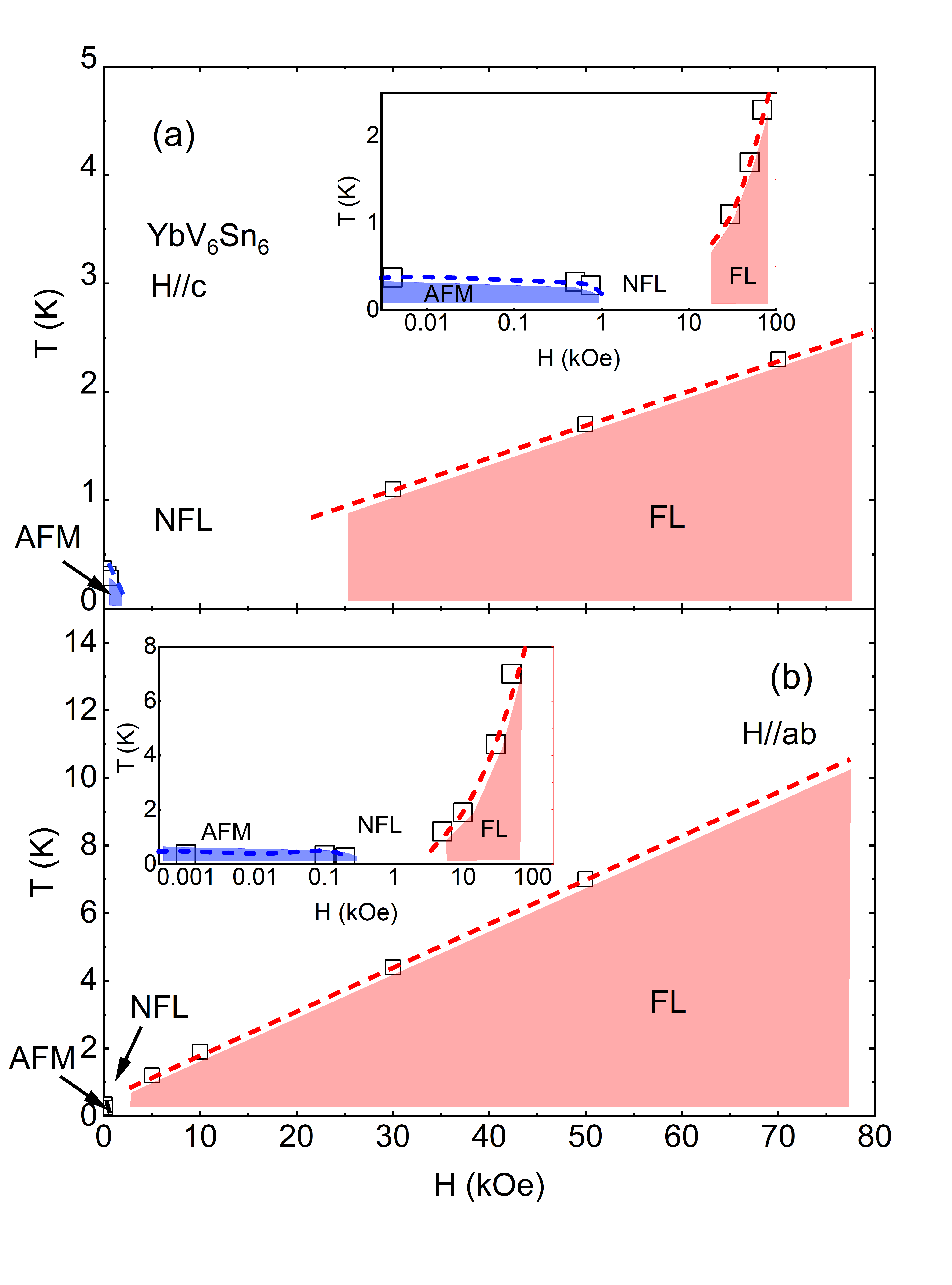}
	\caption{\label{fig8}  A coarse $T-H$ phase diagram of $\mathrm{YbV_6Sn_6}$ when (a), $H\parallel c$ axis; (b), $H\parallel ab$ plane. Insets: $H$ is plotted in a log scale.}
\end{figure}

Based on the $\rho(T)$ data in various magnetic fields, we plot a coarse $T-H$ phase diagram for $\mathrm{YbV_6Sn_6}$  when $H\parallel c$ axis and $H\parallel ab$ plane in Fig.~\ref{fig8}~a and b, respectively.
When the magnetic ordering is suppressed by $H$, we observe a linear dependence of $\rho(T)$, a characteristic feature of an NFL.
However, currently it is not clear whether the NFL region reduces to a QCP or remains a finite range in the $H$ axis at zero temperature.
In other words, we do not know whether the FL region connects with the AFM region at a QCP in 0~K, like the phase diagram in $\mathrm{YbRh_2Si_2}$ \cite{gegenwart2008unconventional}, or detaches from it, like the observed diagrams for $\mathrm{YbAgGe}$ \cite{bud2004magnetic} and Ge-doped $\mathrm{YbRh_2Si_2}$ \cite{custers2003break}.

An interesting observation is the $T_{\mathrm{FL}}$ region shrinks quasi-linearly with decreasing magnetic field in the paramagnetic state.
Using simple linear extrapolation, we find that the  $T_{\mathrm{FL}}$ terminates at $H_{\parallel c}=-6 ~\mathrm{kOe}$ and $H_{\parallel ab}=-3.4~ \mathrm{kOe}$, respectively.
The meaning of the negative termination is not clear because positive termination of $T_{\mathrm{FL}}$ was commonly observed for other HF compounds \cite{donni1996geometrically,mun2013magnetic} .
Further studies on the Hall effect and magnetoresistance will unveil whether and when the Fermi surface is reconstructed in magnetic field.

In the future studies we will understand the nature of the magnetic ordering at $T_{\mathrm{N}}=0.40$~K for $\mathrm{YbV_6Sn_6}$. A PKS state is a plausible ground state for a geometric frustrated Kondo lattice.
On the other hand, a triangular Kondo lattice can also give rise to complex chiral spin structures and even potentially host a skyrmion phase \cite{akagi2010spin,wang2021skyrmion}.
Recent study on $\mathrm{GdV_6Sn_6}$ suggested the existence of a noncollinear spin structure below its $T_{\mathrm{N}}$ \cite{PhysRevB.104.235139}.
The triangular Kondo lattice in $\mathrm{YbV_6Sn_6}$ can also resolve the frustration by forming those unusual magnetic structures in a more local-moment way.
Understanding the magnetic ordering in $\mathrm{YbV_6Sn_6}$  will help to identify its position in the global phase diagram.

\section{Conclusion}

The magnetization, heat capacity and resistivity for the single crystals of $\mathrm{YbV_6Sn_6}$ show typical physical properties of HF compounds while a remarkable magnetic ordering occurs at $0.40$~K.
When the ordering is suppressed by a weak magnetic field, we observed characteristic behaviors of an NFL, including a linear $T$-dependent resistivity,  $-\log(T)/T$ scale of magnetic susceptibility and $-\log(T)$ scale of $C_p/T$.
$\mathrm{YbV_6Sn_6}$ presents as a rare example of the HF compounds bearing Yb-based triangular Kondo lattice, while its magnetic ground state needs further elaboration.
 
\section{ACKNOWLEDGE}
We gratefully thank the discussions with Nanlin Wang, Yuan Li, Yi Zhang, Peijie Sun and H.Q. Yuan.
This work was supported by the National Key Research and Development Program of China (2021YFA1401902), 
the CAS Interdisciplinary Innovation Team, the strategic Priority Research Program of Chinese Academy of Sciences, Grant No. XDB28000000 and the National Natural Science Foundation of China No. 12141002 No.12225401 and No.U1832214.

\bibliography{heavyfermion}% Produces the bibliography via BibTeX.

\end{document}